\documentclass[sigconf, screen, authorversion]{acmart}
\usepackage{pdfpages}
\usepackage{nicefrac}
\usepackage{algorithm}
\usepackage{algpseudocode}
\usepackage{cleveref}
\usepackage{listings}
\usepackage{xcolor}

\definecolor{lightgray}{RGB}{245,245,245}
\definecolor{softblue}{RGB}{70,100,160}

\AtBeginDocument{%
  }

\copyrightyear{2025} 
\acmYear{2025} 
\setcopyright{cc}
\setcctype{by}
\acmConference[SA Technical Communications '25]{SIGGRAPH Asia 2025
Technical Communications}{December 15--18, 2025}{Hong Kong, Hong Kong}
\acmBooktitle{SIGGRAPH Asia 2025 Technical Communications (SA Technical
Communications '25), December 15--18, 2025, Hong Kong, Hong
Kong}\acmDOI{10.1145/3757376.3771378}
\acmISBN{979-8-4007-2136-6/2025/12}

\begin{document}

\title{Geometric Queries on Closed Implicit Surfaces for Walk on Stars}

\author{Tianyu Huang}
\email{tianyu@illumiart.net}
\orcid{0009-0003-5405-4477}
\affiliation{%
  \institution{Independent researcher}
  \city{Beijing}
  \country{China}
}
\affiliation{%
  \institution{Tsinghua University}
  \city{Beijing}
  \country{China}
}

\renewcommand{\shortauthors}{Tianyu Huang}

\begin{abstract}
\emph{Walk on stars (WoSt)} is currently one of the most advanced Monte Carlo solvers for PDEs. Unfortunately, the lack of reliable geometric query approaches has hindered its applicability to boundaries defined by implicit surfaces. This work proposes a geometric query framework over closed implicit surfaces for WoSt, under the scope of \emph{walkin' Robin}~\cite{miller2024robin}. Our key observation is that all WoSt queries can be formulated as constrained global optimization or constraint satisfaction problems. Based on our formulations, to solve the highly non-convex problems, we adopt a branch-and-bound approach based on interval analysis. To the best of our knowledge, our method is the first to study closest silhouette point queries and Robin radius bound queries on closed implicit surfaces. Our formulations and methods first enable mesh-free PDE solving via WoSt when boundaries are defined by closed implicit surfaces.
\end{abstract}


\maketitle

\section{Introduction}

Partial differential equations (PDEs) are fundamental in modeling physical phenomena and widely used in graphics. Recently, Monte Carlo solvers for PDEs have gained increasing attention, with \emph{walk on stars (WoSt)}~\cite{sawhney2023wost, miller2024robin}, an extension of \emph{walk on spheres (WoS)}~\cite{muller1956wos}, emerging as a promising approach. Unlike traditional methods, WoSt avoids discretization and thus bypasses the robustness and performance challenges of volumetric meshing, making it well-suited for increasingly complex geometries in graphics.

Despite these advantages, WoSt has not yet been demonstrated to operate entirely in domains defined by implicit surfaces (hereafter referred to as \emph{implicit domains}). Implicit surfaces can represent intricate geometric details, support robust Boolean/blending operations, and are central to many modern scene representations. A volumetric PDE solver supporting implicit domains directly would inherit these benefits, but current WoSt implementations remain tied to boundary meshes or only support Dirichlet boundaries defined by implicit surfaces~\cite{miller2024diffwos}, limiting its strengths in implicit domains. The difficulty in extending WoSt to implicit domains is that, while WoS only requires closest point queries, WoSt additionally requires geometric queries that have not been well studied on implicit surfaces. To facilitate explanation, we list all the geometric queries required by WoSt:

\begin{enumerate}
    \item Closest point query (CPQ).
    \item Ray intersection query.
    \item Closest silhouette point query (CSPQ).
    \item Robin radius bound query (RRBQ).
    \item Point sampling on \emph{reflecting} (Neumann/Robin) boundaries.
\end{enumerate}

\begin{figure}
    \centering
    \includegraphics[width=0.9\columnwidth]{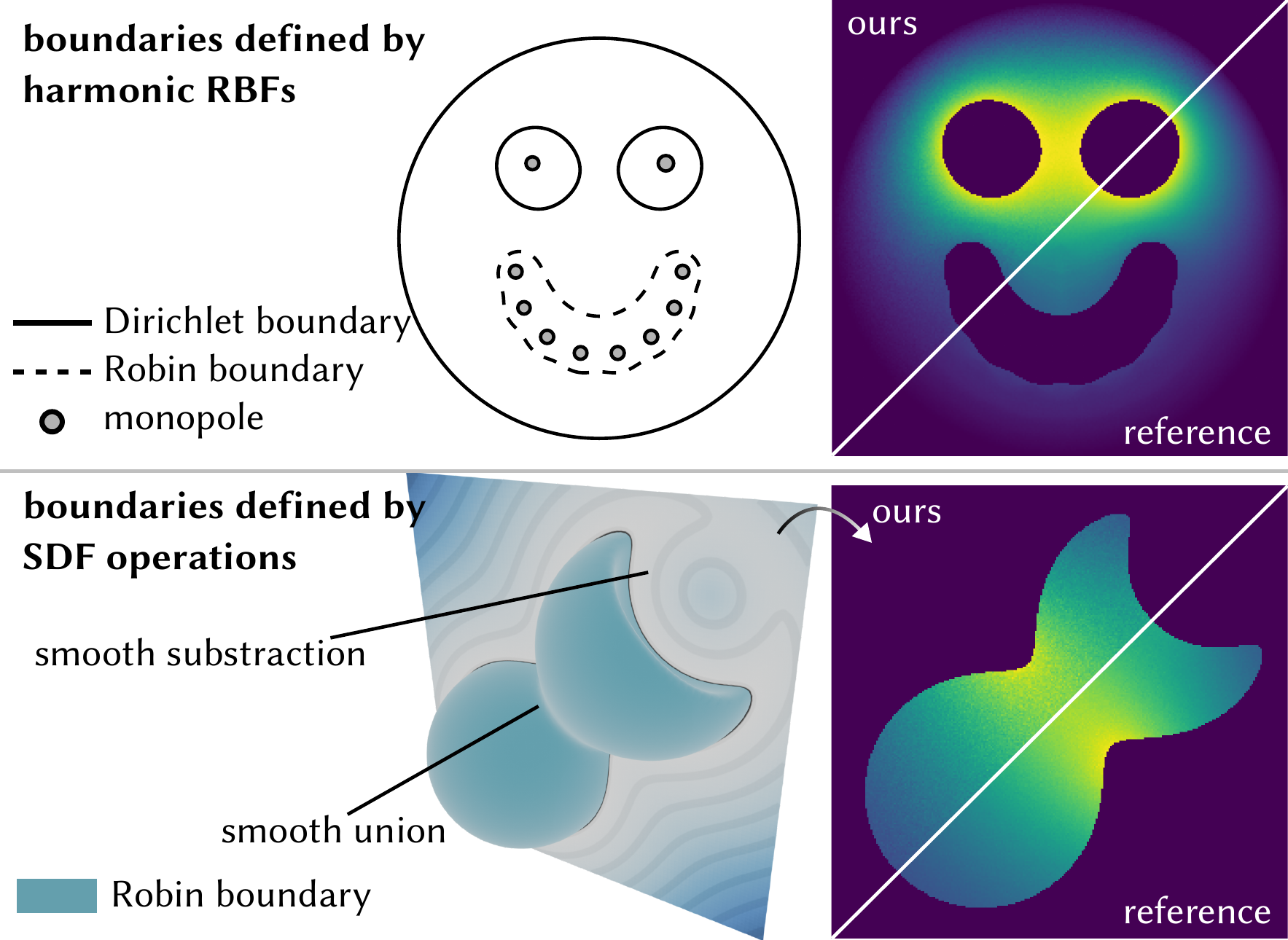}
    \caption{\textbf{Walk on stars (WoSt) with boundaries defined by closed implicit surfaces.} Our method can perform complete WoSt geometric queries on various closed implicit representations (here we demonstrate harmonic RBFs~\cite{miller2024diffwos} and SDF operations). WoSt based on our method yields correct solutions to the Laplace Dirichlet-Robin problems.}
    \label{fig:teaser}
\end{figure}

\noindent
While queries (1), (2), and (5) are well studied, no existing methods address (3) and (4) on implicit surfaces. For instance, silhouette detection has been studied in non-photorealistic~\cite{bremer1998rapid} and differentiable~\cite{zhou2024psdrsdf} rendering, but methods for CSPQ remain unexplored on implicit surfaces; and RRBQ has only been studied for boundary meshes~\cite{miller2024robin}. Consequently, WoSt still falls back to intermediate boundary meshing to handle implicit domains, undermining its mesh-free appeal and degrading geometric fidelity.

In response, we propose a geometric query solution over closed~(compact and boundary-free) implicit surfaces for WoSt, corresponding to PDE problems where the Dirichlet and reflecting boundaries are defined by closed implicit surfaces.
Our approach builds on an observation: all the queries above can be formulated as constrained global optimization or constraint satisfaction problems. Based on our proposed formulations, to run optimization and satisfaction on the non-convex function landscape, we adopt branch-and-bound techniques based on interval analysis (IA)~\cite{snyder1992ia}. From a contributions perspective, we are the first to study the formulation and implementation of queries (3) and (4) on closed implicit surfaces, and the first to realize a complete WoSt when boundaries are defined by closed implicit surfaces.

\section{Preliminary: Interval Analysis (IA) for Constrained Global Optimization and Constraint Satisfaction}

\label{sec:prel}

The geometric queries exhibit non-convex landscapes, and WoSt often admits conservative queries. Consequently, we adopt IA as the tool for our queries. We recall two relevant methods introduced by \citet{snyder1992ia}: \texttt{MINIMIZE} for constrained global optimization, and \texttt{SOLVE} for constraint satisfaction. In the methods, \emph{boxes}, \textit{i.e.}, Cartesian product of intervals, are used to divide the space. 
For an \emph{objective function} $g$, its corresponding \emph{objective inclusion function} $\Box g$ maps a box $Y$ to an interval enclosing all $g(x), \forall x \in Y$. 
\emph{Constraint inclusion function} $\Box G$ that takes a box $Y$ returns a three-value logic:
\begin{itemize}
  \item \texttt{positive}: constraint $G$ holds for all points in $Y$,
  \item \texttt{negative}: constraint $G$ fails for all points in $Y$,
  \item \texttt{unknown}: both satisfying and unsatisfying points exist in $Y$.
\end{itemize}
\noindent
\emph{Solution acceptance set constraint} $\Box A$ returns \texttt{true} if the search box is fine enough (that meets the specified tolerance). \texttt{X} is the initial box. The algorithms are as follows:

\begin{lstlisting}[caption={\texttt{MINIMIZE}: Constrained Global Optimization}, label={alg:minimize}, language=Python]
def MINIMIZE( $\Box g$, $\Box G$, $\Box A$, X ):
    L $\leftarrow$ PriorityQueue({X})
    UB $\leftarrow +\infty$ # upper bound for pruning
    results $\leftarrow \varnothing$ # multiple solutions may arise
    while L $\neq \varnothing$:
        Y $\leftarrow$ L.pop_min_lower_bound() # best candidate
        if $\Box A$(Y) = true: # stricter tests may be applied
            results $\leftarrow$ results $\cup$ {Y} # accept Y
        else:
            for Y_i in subdivide(Y):
                if $\Box G$(Y_i) = negative:
                    continue # prune constraint-violating boxes
                [a_i, b_i] $\leftarrow$ $\Box g$(Y_i)
                if a_i > UB:
                    continue # prune suboptimal boxes
                L.insert(Y_i, priority=a_i)
                UB $\leftarrow $ min(UB, b_i) # update upper bound
    return results
\end{lstlisting}

\begin{lstlisting}[caption={\texttt{SOLVE}: Constraint Satisfaction}, label={alg:solve}, language=Python]
def SOLVE( $\Box G$, $\Box A$, X ):
    L $\leftarrow$ {X}, results $\leftarrow \varnothing$
    while L $\neq \varnothing$:
        Y $\leftarrow$ remove(L) # take a box from L
        if $\Box G$(Y) = positive $\land$ $\Box A$(Y) = true:
            results $\leftarrow$ results $\cup$ {Y} # accept fine & constraint-satisfying boxes
        elif $\Box G$(Y) = unknown:
            L $\leftarrow$ L $\cup$ subdivide(Y) # further subdivide unknown boxes
    return results
\end{lstlisting}

In our paper, \texttt{SOLVE} is used for point sampling on reflecting boundaries (\cref{sec:point_sampling}), while the rest utilizes \texttt{MINIMIZE}.

\section{Method}

We consider a domain $\Omega \subset \mathbb{R}^d$ $(d=2,3)$ whose boundary\footnote{In this section, "boundary" refers to the boundary of the PDE problems and should be distinguished from the boundaries of the implicit surfaces discussed in the context of open and closed surfaces.} is partitioned into two closed implicit surfaces: the Dirichlet boundary $\partial\Omega_\text{D}$ and the reflecting boundary $\partial\Omega_\text{R}$. They are defined as the zero level sets of smooth implicit functions $f_\text{D}, f_\text{R}:\mathbb{R}^d \to \mathbb{R}$, with the regularity condition $\nabla f_\text{D}(x)\neq 0, \forall x\in\partial\Omega_\text{D}$ and $\nabla f_\text{R}(z)\neq 0, \forall z\in\partial\Omega_\text{R}$. Under this condition, $\nabla f_\text{D}$ and $\nabla f_\text{R}$ are parallel to the corresponding surface normals. No Lipschitz assumption is required. We denote by $x \in \Omega$ a query point, by $y \in \partial\Omega_\text{D}$ a Dirichlet boundary point, and by $z \in \partial\Omega_\text{R}$ a reflecting boundary point. The closed and open balls centered at $x$ with radius $R$ are written as $B(x,R)$ and $\mathring{B}(x,R)=B(x,R)\setminus\partial B(x,R)$, \textit{resp.} Some notations here differ from those in the WoSt paper for contextual consistency; readers unfamiliar with WoSt may refer to the supplementary material.

\subsection{Closest Point Query (CPQ) \& Ray Intersection Query}

In WoSt, CPQ is used on $\partial\Omega_\text{D}$ and ray intersection query is used on $\partial\Omega_\text{R}$, these queries are well-studied as global optimization problems; for completeness, we briefly summarize them here. The CPQ can be formulated as:

\begin{equation}
y_\text{D} = \arg\min_{y \in \mathbb{R}^d} \|y - x\|^2 \quad \text{subject to} \quad f_\text{D}(y) = 0,
\end{equation}

\noindent
and ray intersection query can be formulated as:

\begin{equation}
z_\text{I} = \arg\min_{z \in \mathbb{R}^d} \|z - x\|^2 \quad \text{subject to} \, f_\text{R}(z) = 0,\; z \in \{ x + t v \mid t > 0 \},
\end{equation}

\noindent
where $v \in \mathbb{R}^d$ is a unit direction vector. Apart from using \texttt{MINIMIZE}, many efficient methods exist, which we do not discuss further here. CPQ yields the \emph{Dirichlet distance} $R_\text{D} = \|y_\text{D}-x\|$.

\subsection{Closest Silhouette Point Query (CSPQ)}

\label{sec:cspq}

A silhouette point is a point whose normal is perpendicular to the viewing direction. In WoSt, CSPQ is used on $\partial\Omega_\text{R}$ to produce the \emph{silhouette bound} $R_\text{S}$ for the radius of the star-shaped region such that no silhouette exists within the star-shaped region. We propose to formulate CSPQ as the following problem when $x\notin\partial\Omega_\text{R}$:

\begin{equation}
z_\text{S} = \arg\min_{z \in \mathbb{R}^d} \|z - x\|^2 \quad \text{subject to} \quad 
\begin{cases}
f_\text{R}(z) = 0 \\
\nabla f_\text{R}(z) \cdot (z-x) = 0,
\end{cases}
\end{equation}

\noindent
therefore we have $R_\text{S}=\|z_\text{S}-x\|$. In WoSt, $R_\text{S}$ should satisfy $R_\text{S} \leq R_\text{D}$, thus we introduce the additional constraint $\|z - x\| \leq R_\text{D}$. To narrow the search space, we initialize the \texttt{MINIMIZE} algorithm with a box centered at $x$ and of side length $2R_\text{D}$. When the problem involves Robin boundaries, we slightly shrink $R_\text{S}$ to ensure $\cos \theta(z) = \frac{n_z \cdot (z - x)}{r} = \frac{\left|\nabla f_\text{R}(z) \cdot (z - x)\right|}{\|\nabla f_\text{R}(z)\| \, r}>0, \forall z \in B(x,R_\text{S})\cap \partial\Omega_\text{R}$\footnote{$\nabla f_\text{R}$ always points outward from the implicit surface, irrespective of $x$'s location (inside or outside); thus we take the absolute value in the dot product.}, thereby avoiding singularities in \emph{reflectance} $\rho_\mu$~\cite[eqs. 9 and 33]{miller2024robin}. The query strategy for $x\in\partial\Omega_\text{R}$ is discussed in the supplementary material.

\begin{figure}
    \centering
    \includegraphics[width=0.9\columnwidth]{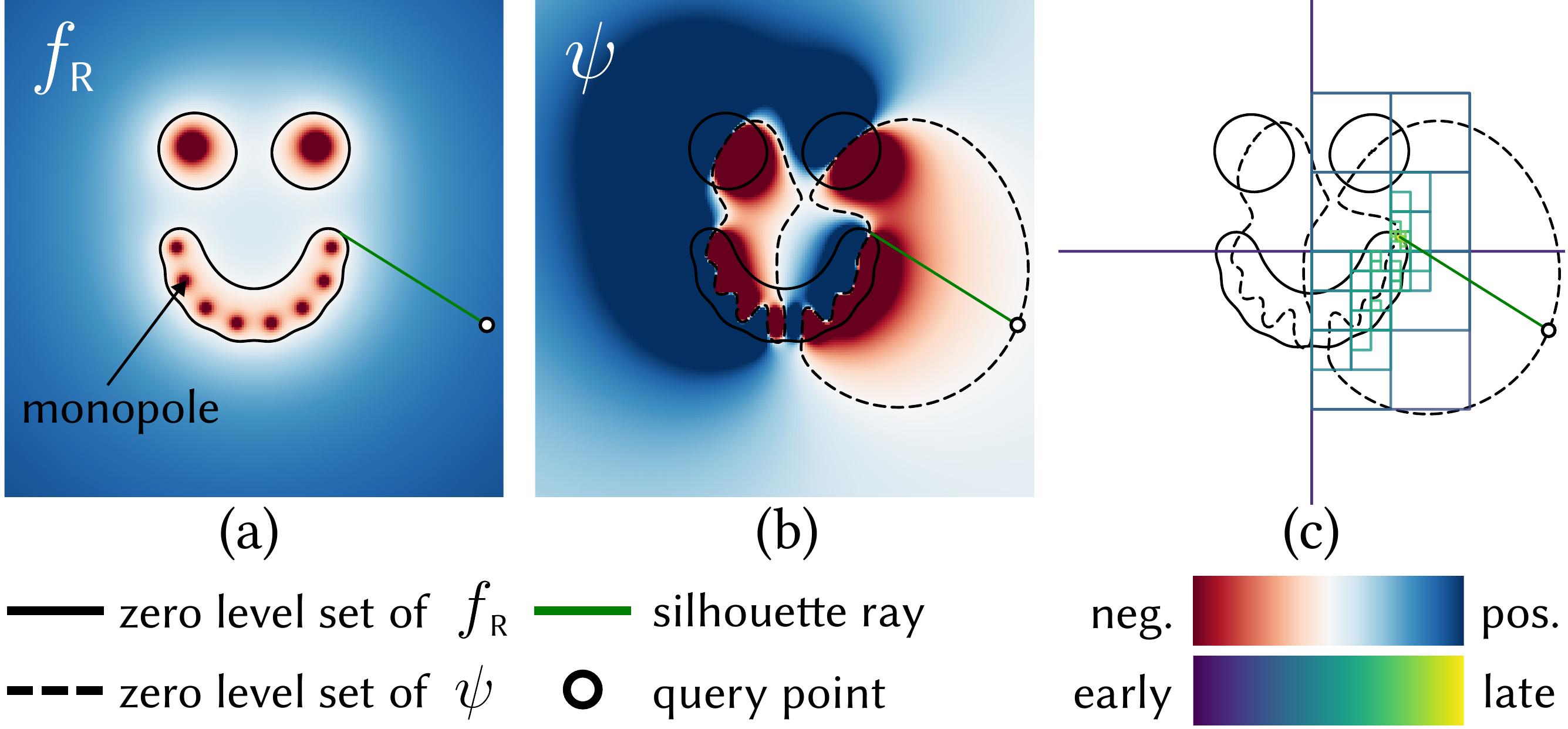}
    \caption{\textbf{Closest silhouette point query (CSPQ).} Figures (a) and (b) illustrate the landscapes of $f_\text{R}$ and $\psi = \nabla f_\text{R}(z) \cdot (z-x)$, \textit{resp.} Figure (c) visualizes the minimization process.}
    \label{fig:closest_silhohuette_query}
\end{figure}

\subsection{Robin Radius Bound Query (RRBQ)}
\label{sec:rrbq}
In walkin' Robin, to ensure $\rho_\mu\in[0, 1]$ for a stable Monte Carlo estimator, \citet{miller2024robin} propose shrinking $R_\text{S}$ to the \emph{Robin radius bound} $R_\text{R}\leq R_\text{S}$. We now discuss the query for $R_\text{R}$ by dimension.

\subsubsection{2D Case}
In 2D, a desired $R_\text{R}$ ensures, $\forall z \in B(x, R_\text{R})\cap \partial\Omega_\text{R}\subset B(x, R_\text{S})\cap \partial\Omega_\text{R}$, the following holds~\citep[eq. 35]{miller2024robin}:

\begin{equation}
\label{eq:2d_constraint}
R_\text{R} \leq r \exp \left(\frac{\cos \theta}{\mu(z) \, r} \right),
\end{equation}

\noindent
where $r = r(z) = \lVert z-x \rVert$, $\cos \theta>0$ (\cref{sec:cspq}), and $\mu(z) > 0$ is the prescribed Robin coefficient defined on $\partial\Omega_\text{R}$. 
Since $z$ appears entirely on the right-hand side of \cref{eq:2d_constraint}, it is natural to consider taking the minimum of it over $B(x,R_\text{S})\cap\partial\Omega_\text{R}$ as the upper bound for $R_\text{R}$, which corresponds to the following optimization problem:

\begin{equation}
\label{eq:2d_robin_opt}
R_\text{R}^* = \min_{z \in \mathbb{R}^2} \ r \exp \left(\frac{\cos \theta}{\mu(z) \, r} \right) \quad \text{subject to} \quad 
\begin{cases}
f_\text{R}(z) = 0 \\
r \leq R_\text{S}.
\end{cases}
\end{equation}

\noindent
We further find that the bound $R_\text{R}^*$ here is tight, \textit{i.e.}, there exists no $\tilde{R_\text{R}} > R_\text{R}^*$ satisfying \cref{eq:2d_constraint}, because $\forall z \in B(x,R_\text{S})\cap \partial\Omega_\text{R}$ we have:

\begin{equation}
\label{eq:good_2d}
r \exp \left(\frac{\cos \theta}{\mu(z) \, r} \right) - r = r \left[\exp \left(\frac{\cos \theta}{\mu(z) \, r} \right)-1\right]>0.
\end{equation}

\noindent
Therefore, the minimizer $z^*$ necessarily satisfies $r^*(z^*) < R_\text{R}^*$, or equivalently $z^* \in \mathring{B}(x, R_\text{R}^*) \cap \partial\Omega_\text{R}$, so $\tilde{R_\text{R}}>R_\text{R}^*$ would necessarily violate \cref{eq:2d_constraint} at $z^*$. Hence, $R_\text{R}^*$ is guaranteed to be the tight bound.

Finally, we take the lower bound of the interval for $R_\text{R}^*$ for numerical stability. This also applies to the 3D case.

\begin{figure}
    \centering
    \includegraphics[width=0.9\columnwidth]{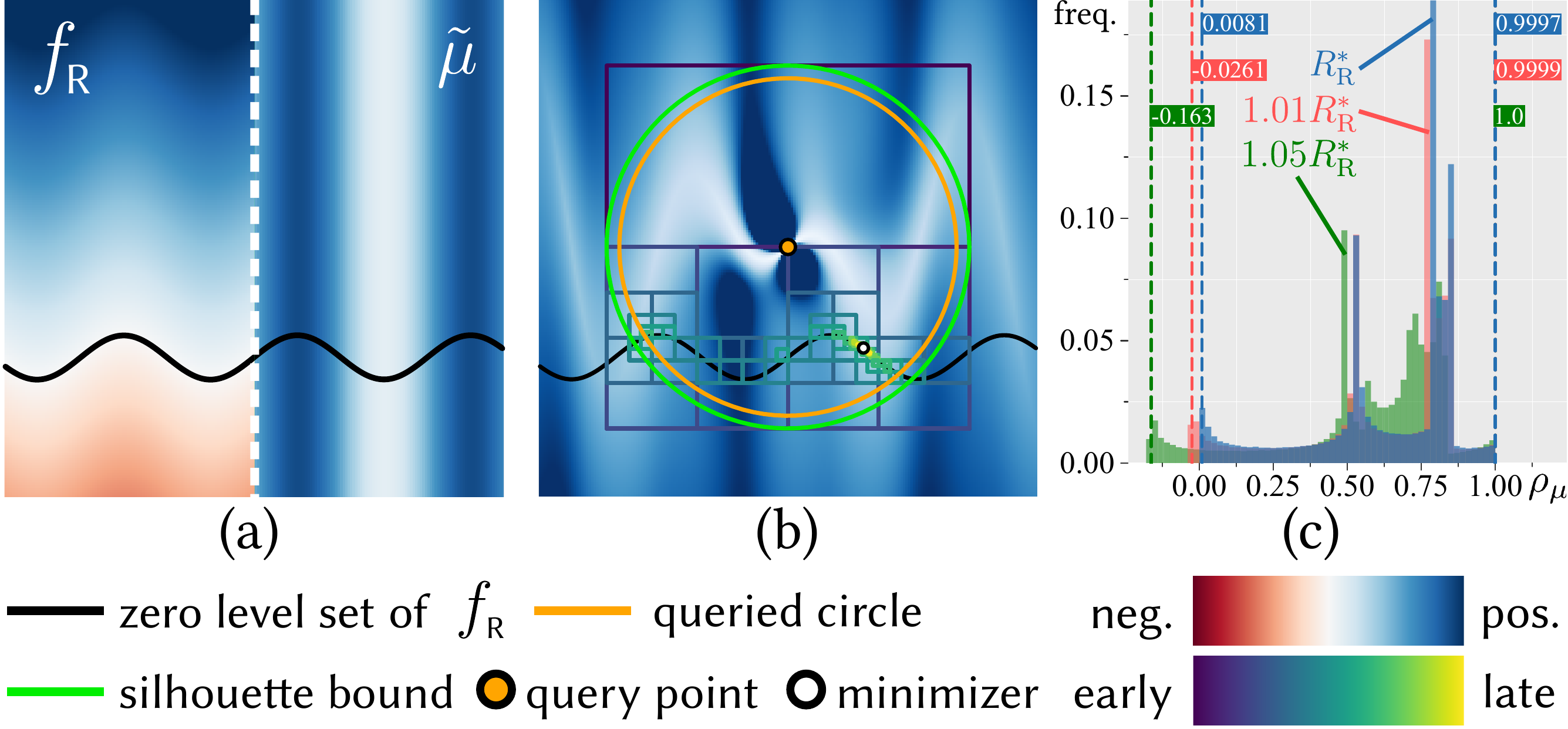}
    \caption{\textbf{Robin radius bound query (RRBQ).} Figure (a) shows the landscape of $f_\text{R}$ and $\tilde{\mu}$ (\cref{sec:rrbq_disc}). Figure (b) shows the landscape of the objective, optimization process and result. In figure (c), we randomly sample points on $\partial\Omega_\text{R}$ inside the queried circle, and plot the distribution of $\rho_\mu$~\cite[eqs. 9 and 33]{miller2024robin}. As illustrated, $\rho_\mu \in [0, 1]$ when the radius equals to $R_\text{R}^*$, while increasing the radius slightly yields invalid ranges of $\rho_\mu$.}
    \label{fig:robin_radius_query}
\end{figure}

\begin{figure}
    \centering
    \includegraphics[width=0.9\columnwidth]{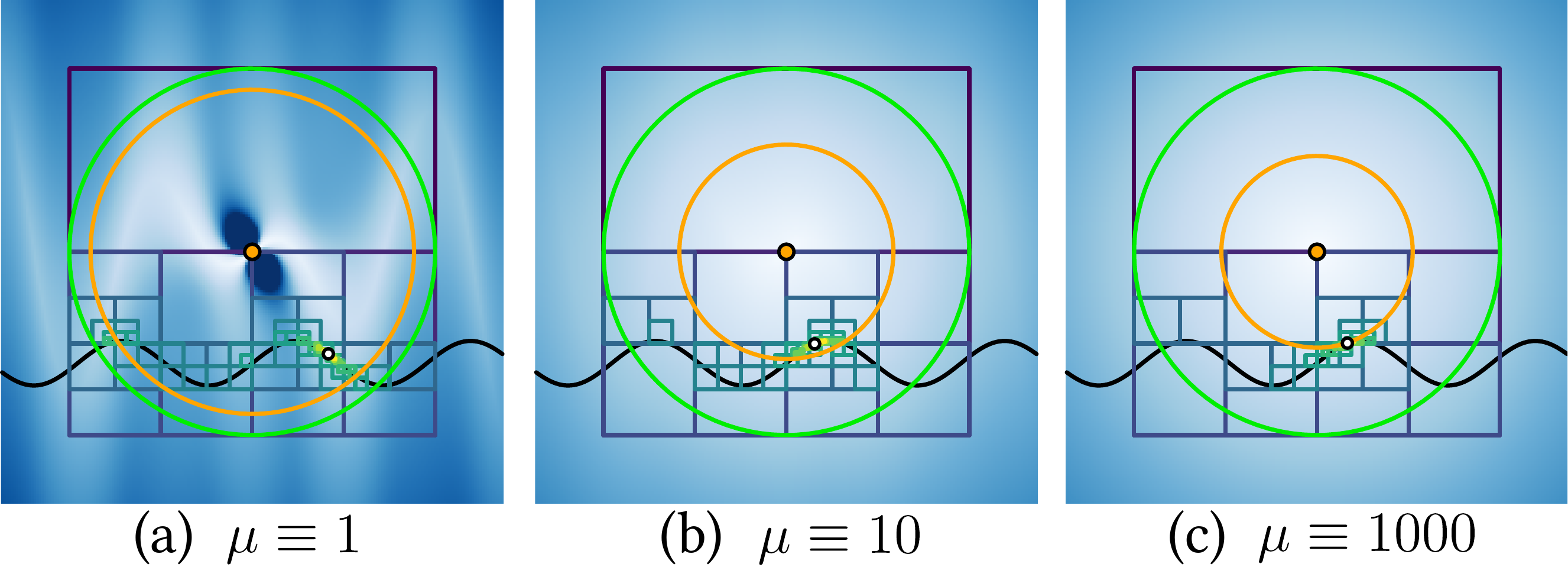}
    \caption{\textbf{Sanity check on the effect of increasing $\mu$ on the radius.} We set $\tilde{\mu}(z)$ to a constant \textit{w.r.t.} $z$, increase it from (a) to (c), and perform RRBQ. The results show the query converges to CPQ on $\partial\Omega_\text{R}$ as $\mu$ increases, which agrees with the results reported by \citet[fig. 7]{miller2024robin}.}
    \label{fig:robin_empirical}
\end{figure}

\subsubsection{3D Case.} 

In 3D, \cref{eq:2d_constraint} takes a different form~\cite[eq. 11]{miller2024robin}:

\begin{equation}
\label{eq:3d_constraint}
R_\text{R} \leq \frac{r}{1 - \frac{\cos \theta}{\mu(z) r}} \qquad \text{when} \quad r > \frac{\cos \theta}{\mu(z)}.
\end{equation}

\noindent
The definition of symbols follow the 2D case except that the points and vectors are defined in $\mathbb{R}^3$. The biggest difference lies in the condition $r>\frac{\cos\theta}{\mu(z)}$. When $0 < r \leq \frac{\cos \theta}{\mu(z)}$, $R_\text{R}$ is no longer subject to \cref{eq:3d_constraint}, and it can be made arbitrarily large~\cite[section 4.2.1]{miller2024robin}.
We also observe that \cref{eq:3d_constraint} naturally guarantees:

\begin{equation}
\label{eq:good_3d}
\frac{r}{1 - \frac{\cos \theta}{\mu(z) r}} - r = \frac{\frac{\cos \theta}{\mu(z)}}{1 - \frac{\cos \theta}{\mu(z) r}} > 0 \qquad \text{when} \quad r > \frac{\cos \theta}{\mu(z)}.
\end{equation}

\noindent
Thus, the following optimization problem also yields a tight bound:

\begin{equation}
\label{eq:opt_problem_3d}
R_\text{R}^* = \min_{z \in \mathbb{R}^3} \ \frac{r}{1 - \frac{\cos \theta}{\mu(z) \, r}} \quad \text{subject to} \quad
\begin{cases}
f_\text{R}(z) = 0 \\
\frac{\cos \theta}{\mu(z)} < r \leq R_\text{S}.
\end{cases}
\end{equation}

\noindent
We denote the objective function in \cref{eq:opt_problem_3d} as $O(z;x)$. Using IA to impose $r > \nicefrac{\cos\theta}{\mu(z)}$ may be overkill, as in condition-violating regions, we no longer need to evaluate its range. We instead adopt a simpler approach: based on the observation that $O(z; x) \to +\infty$ as
$r \to \left(\nicefrac{\cos\theta}{\mu(z)}\right)^{+}$, we set a new optimization objective:

\begin{equation}
\label{eq:new_3d_optimizer}
\tilde{O}(z; x)=
\begin{cases}
    \frac{r}{1 - \frac{\cos \theta}{\mu(z) \, r}}, \qquad & r > \frac{\cos \theta}{\mu(z)} \\
    +\infty, \qquad & \textit{o.w.}
\end{cases}
\end{equation}

\noindent
Since we are performing a minimization, the optimizer will automatically bypass the $+\infty$ region.
This objective can be computed numerically by clamping the interval of $1 - \frac{\cos \theta(z)}{\mu(z) \, r}$ to $[0, +\infty)$ (we define division by 0 to yield $+\infty$). In the end we have:

\begin{equation}
R_\text{R}^* = \min_{z \in \mathbb{R}^3} \ \tilde{O}(z;x) \quad \text{subject to} \quad
\begin{cases}
f_\text{R}(z) = 0 \\
r \leq R_\text{S}.
\end{cases}
\end{equation}

\begin{figure}
    \centering
    \includegraphics[width=0.9\columnwidth]{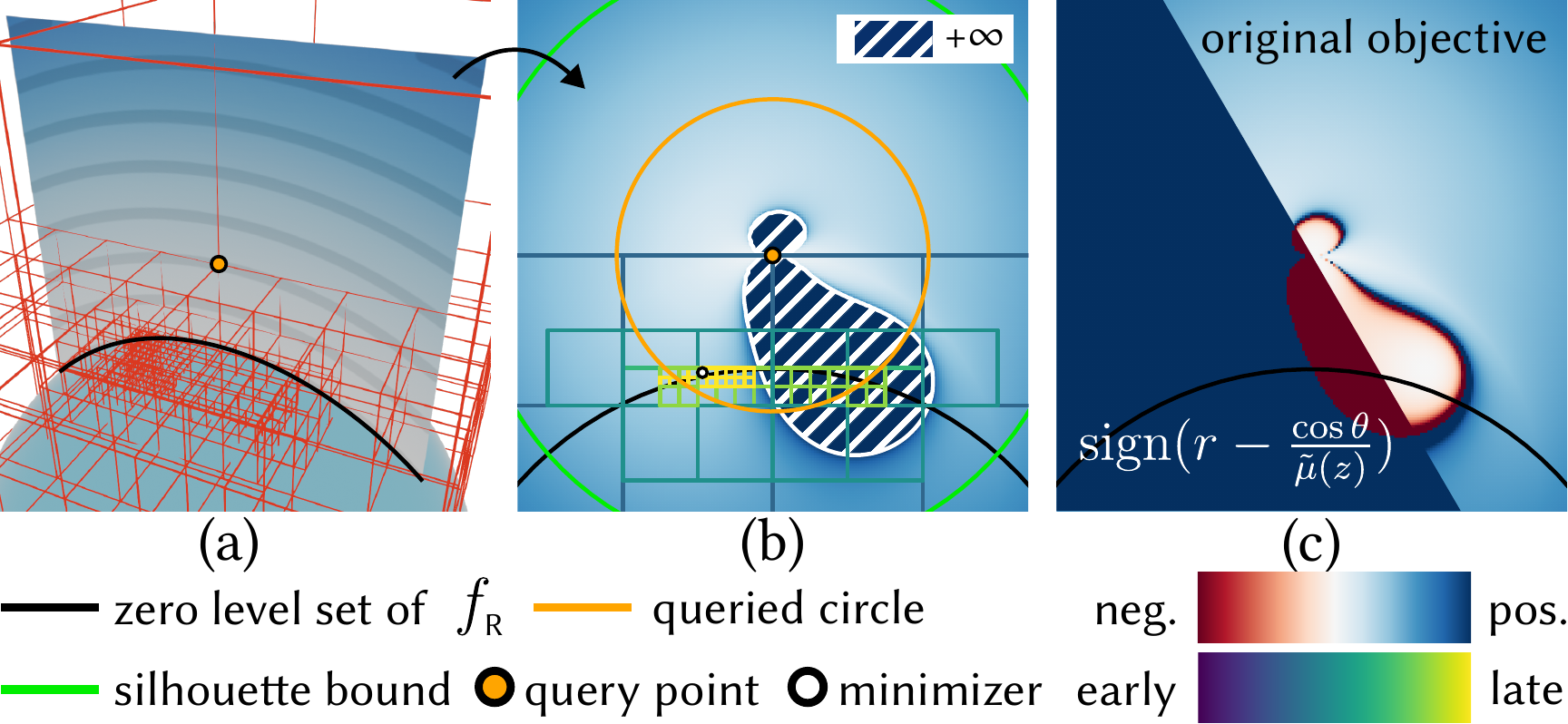}
    \caption{\textbf{RRBQ in 3D.} Figures (a) and (b) show the optimization process and results in $\mathbb{R}^3$ and on a slice, \textit{resp.} Comparing (b) and (c), we see that regions where $r>\frac{\cos\theta}{\tilde{\mu}(z)}$ is not satisfied are set to $+\infty$ in \cref{eq:new_3d_optimizer}, then the optimizer skips these regions during the coarse stage. The sampled values (see \cref{fig:robin_radius_query} caption) of $\rho_\mu$ fall within $[0.4891, 0.99996]$.}
    \label{fig:robin_radius_3d}
\end{figure}

\subsubsection{Discussion \& Implementation}
\label{sec:rrbq_disc}
By definition, the domain of $\mu$ is $\partial\Omega_\text{R}$, whereas the IA operates in $\mathbb{R}^d$. Therefore, we extend $\mu$ to $\mathbb{R}^d$ by constructing $\tilde{\mu}: \mathbb{R}^d \to \mathbb{R}$ that satisfies $\tilde{\mu}(z) = \mu(z), \forall z \in \partial\Omega_\text{R}$. We initialize the \texttt{MINIMIZE} algorithm with a box centered at $x$ with side length $2R_\text{S}$.

\subsection{Point Sampling on Reflecting Boundaries}

\label{sec:point_sampling}

In WoSt, we need to sample points on $\Gamma =B(x,R_\text{R}) \cap \partial\Omega_\text{R}$~\cite[section 5.2]{sawhney2023wost}, which can be formalized as the following constraint satisfaction problem:

\begin{equation}
\Gamma := \{ z \in \mathbb{R}^d \mid f_\text{R}(z) = 0,\ \|z - x\| \leq R_\text{R} \}.
\end{equation}

\noindent
We adopt the \texttt{SOLVE} algorithm. The initial search domain is the box centered at $x$ and of side length $2R_\text{R}$. In practice, \emph{reservoir sampling} or a cached tree structure can be used to improve performance.

\begin{figure}
    \centering
    \includegraphics[width=0.9\columnwidth]{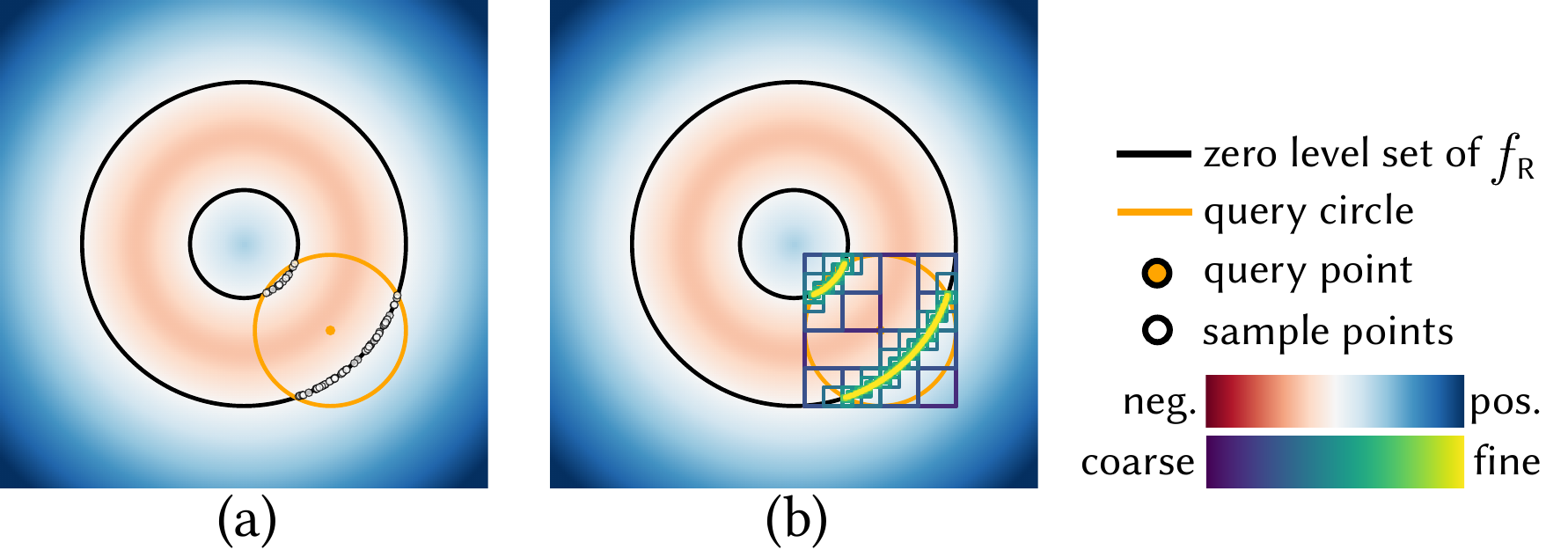}
    \caption{\textbf{Point sampling on reflecting boundaries.} Figure (a) shows the implicit function landscape, the query circle, and the sampled points. Figure (b) illustrates the search boxes.}
    \label{fig:robin_boundary_sampling}
\end{figure}

\section{Results and Discussion}

We implement the method in Julia~\cite{julia2017} using the \texttt{IntervalArithmetic.jl}~\cite{sanders2014ia} library, extending it with forward-mode automatic differentiation via dual intervals to obtain gradients. We set the tolerance in \cref{sec:prel} to $\nicefrac{1}{10}$ of the $\epsilon$-shell.
For each newly introduced query, we conduct verification experiments in \cref{fig:closest_silhohuette_query,fig:robin_radius_query,fig:robin_empirical,fig:robin_radius_3d,fig:robin_boundary_sampling}. Based on our method, we also use WoSt to solve Laplace equations, as shown in \cref{fig:teaser}. The results match the expected behavior. We provide detailed setup in the supplementary document.

\section{Conclusion, Limitations and Future Work}

By proposing formulations and adopting IA-based global optimization and constraint satisfaction methods, our present method first enables all necessary geometric queries on closed implicit surfaces for WoSt, enabling mesh-free PDE solving via WoSt in implicit domains. Our work comes with limitations: 
although our method is, in principle, applicable to neural SDFs~\cite{sharp2022ra}, we have not implemented this extension; our current boundary sampling strategy is relatively inefficient---recent advances~\cite{ling2025uniform} may be combined to improve performance; our work primarily serves as a proof-of-concept, leaving high-performance implementations for future work. At present, optimizing implicit surfaces with reflecting boundaries remains an open inverse problem in the Monte Carlo for PDEs community, and we believe our method could serve as a tool toward this goal.

\begin{acks}
Tianyu Huang would like to thank Hao Pan, and Ryusuke Sugimoto for valuable discussions. Jingwang Ling helped with proofreading of this paper. This paper utilized resources from Feng Xu's lab.
Tianyu Huang also acknowledges the travel funding support from the Tsinghua University \emph{Spark} Program.
\end{acks}

\bibliographystyle{ACM-Reference-Format}
\bibliography{sample-base}



\section*{Supplementary material}

## A  Notes on the Mathematical Notation in Walk on Stars (WoSt)

For contextual consistency, we use specific symbols and names for certain WoSt concepts. To help readers unfamiliar with WoSt relate this work to the WoSt paper, the following table maps our notation to that of the walkin' Robin paper [Miller et al. 2024]:

Table 1: Mapping of notation between our work and Miller et al. [2024]

| Our work | Miller et al. [2024] |
| --- | --- |
| Dirichlet distance $R_D$ | distance to the closest point $R_\infty = d_{\text{Dirichlet}}(x)$ |
| silhouette bound $R_S$ | distance to the closest silhouette point $R_0 = d_{\text{silhouette}}(x)$ |
| Robin radius bound $R_R$ | $R_\mu$, referred to as $R$ in Miller et al. [2024, section 4.2] |

## B  Notes on $R_S$ Shrinking in CSPQ

In CSPQ (section 3.2), we propose slightly shrinking $R_S$ to avoid singularities in $\rho_\mu$, primarily due to the use of the $\rho_\mu$ notation from Miller et al. [2024, eqs. 7, 9 and 33], namely:

$$\rho_\mu(x, z) = 1 - \mu(z)\frac{G^B(x, z)}{P^B(x, z)}, \qquad (1)$$

where $P^B(x, z)$, the Poisson kernel, is proportional to the projected solid angle; that is, when $\cos\theta(z) = 0$, $P^B(x, z) = 0$, resulting in a singularity. We refer to Sawhney et al. [2023, appendix A] for detailed expression of $P^B$ and the Green's function $G^B$. In the original walkin' Robin work [Miller et al. 2024], this issue is not discussed because they operate on meshes, which have discretely varying normals that automatically ensure $\cos\theta > 0$ within $B(x, R_S)$; while in our work, we consider smooth implicit surfaces, so silhouette points with $\cos\theta = 0$ are included in $B(x, R_S)$.

However, we note that our shrinking operation is primarily for the convenience of subsequent discussion in the main text. Theoretically, even without shrinking, the stability of the sampler is unaffected (though efficiency may be impacted). This is mainly because, in the walkin' Robin estimator [Miller et al. 2024, eq. 8], the recursive term performs *importance sampling over the Poisson kernel*, so silhouettes with $\cos\theta = 0$ have zero probability of being sampled. Even without using the walkin' Robin's default sampler, for a good sampler that attempts to satisfy the optimal sampling probability [Miller et al. 2024, eq. 8]:

$$p \propto \frac{\rho_\mu(x_k, x_{k+1})P^B(x_k, x_{k+1})u(x_{k+1})}{\alpha(x_k)}, \qquad (2)$$

we observe that there is also a $P^B$ factor in the numerator, which cancels the singularity. Furthermore, since $G^B = 0$ on $\partial B$, the contribution and ideal sampling probability at the silhouette points are also zero.

## C  CSPQ on $\partial\Omega_R$

When $x \in \partial\Omega_R$, the optimization problem in the main text would always yield $z_S = x$, and thus $R_S = 0$. This outcome is *acceptable* because, in WoSt, the radius is clamped to a small $\epsilon$, and the walk is still able to proceed. In fact, in walkin' Robin this is a standard strategy to handle cases where $x \in \partial\Omega_R$ but $\mu(x) \neq 0$ [Sawhney and Miller 2023]. There are ways to address this issue, but it is worth noting that simply discarding the current point is not feasible: when the local geometry of $\partial\Omega_R$ at $x$ is concave, the silhouette may indeed lie exactly at $x$. Fortunately, determining this geometric property is not difficult: in systems with second-order derivatives, the Hessian can be used, and otherwise one may resort to finite-difference methods. Inspired by common tricks in ray tracing, one may also consider applying a slight offset along the normal direction to query points that lie on the boundary.

In our implementation, we follow the principle of *Occam's razor* and apply no special handling, except for an $\epsilon$-clamp. A more robust solution will be presented in our future full work.

## D  Discussion of the Case $\mu(z) = 0$

In section 3.3, we assume $\mu(z) > 0$, mainly to avoid singularities in eqs. (5) and (7) in the main text. However, when $\mu(z) = 0$, we have $\rho_\mu = 1 \in [0, 1], \forall z$ [Miller et al. 2024, eqs. 9 and 33]. Thus, similar to the treatment in eq. (10) in the main text, we may regard $R_R$ as arbitrarily large at points $z$ where $\mu(z) = 0$.

## E  Experimental Setup

### E.1  Forward Automatic Differentiation (AD) Implementation

For the proof-of-concept experiments, we did not implement a sophisticated AD system; instead, we used a simple forward-mode AD system based on dual intervals. We employed a `IntervalDual` structure to store both an interval and its gradient:

```
struct IntervalDual <: Real
    val::Interval
    der::Vector{Interval} # "dual" gradient
end
```

For each mathematical operator, we overload a `IntervalDual` version; for example, for division:

```
1  /(a::IntervalDual, b::IntervalDual) = IntervalDual(
2      a.val / b.val,
3      [(a_i * b.val - a.val * b_i) / (b.val^2) for (a_i,
           b_i) in zip(a.der, b.der)] # gradient arithmetic
4  )
```

This allows us to seamlessly switch to `IntervalDual` arithmetic just like ordinary mathematical operations. For example, in the following harmonic RBF boundary function, the implementation is essentially identical to standard mathematical notation:

```
1  function harmonic_rbf(v)
2      # ...parameter definition...
3      h = c
4      for n in 1:N
5          dist = sqrt((x - p[n][1])^2 + (y - p[n][2])^2)
6          h += a[n] / dist
7      end
8      return h
9  end
```

A similar idea has appeared in recent interval analysis work; for instance, Sharp and Jacobson [2022] replace the standard floating point arithmetic in the forward pass of an MLP with affine arithmetic (a tighter form of interval arithmetic). In principle, our method could also be applied to forward or backward derivatives of MLPs, but this would require substantial engineering effort, which we leave for future work.

Using a non-vectorized system on the CPU results in relatively low performance in our implementation. However, we believe that building a high-performance implementation on top of our work would not pose significant challenges, and the branch-and-bound algorithm is inherently highly parallelizable.

### E.2 PDE Solution Generation

For reference solution, we extract the implicit surface using the marching cubes algorithm [Lorensen and Cline 1987] with a resolution of $256^d$ ($d = 2, 3$), and compute the reference solution with 8192 wpp using `Zombie` [Sawhney and Miller 2023]. Our method is limited by the performance of the symbolic computation system, so we generate experimental results with less wpp. The 2D results were computed with 1024 wpp, while the 3D results used 256 wpp. When sampling the boundary, we follow the approach of Sharp and Jacobson [2022] by first constructing and caching a tree structure to improve performance. In one set of experiments, the evaluation points for different geometric representations (meshes and implicit surfaces) were generated using the same mask. This ensures that the solutions produced by different representations yield identical contours on the slice, eliminating contour inconsistencies caused by small floating-point errors and facilitating side-by-side comparison.



\end{document}